\documentclass[prd,twocolumn,superscriptaddress,altaffilletter,amssymb,showpacs,epsfig,nofootinbib]{revtex4}
\usepackage[dvips]{graphicx}
\usepackage{amsmath}
\newcommand{\be}{\begin{equation}}
\newcommand{\ee}{\end{equation}}
\newcommand{\ben}{\begin{eqnarray}}
\newcommand{\een}{\end{eqnarray}}

\def\be{\begin{equation}}
\def\ee{\end{equation}}
\def\bea{\begin{eqnarray}}
\def\eea{\end{eqnarray}}

\begin{document}

\title{Dynamics of Quintessence Models of Dark Energy with Exponential
Coupling to the Dark Matter}
\author{Tame Gonzalez, Genly Leon and Israel Quiros}
\affiliation{Universidad Central de Las Villas, Santa Clara CP
54830, Cuba}
\date{\today}

\begin{abstract}
We explore quintessence models of dark energy which exhibit
non-minimal coupling between the dark matter and the dark energy
components of the cosmic fluid. The kind of coupling chosen is
inspired in scalar-tensor theories of gravity. We impose a suitable
dynamics of the expansion allowing to derive exact
Friedmann-Robertson-Walker solutions once the coupling function is
given as input. Self-interaction potentials of single and double
exponential types emerge as result of our choice of the coupling
function. The stability and existence of the solutions is discussed
in some detail. Although, in general, models with appropriated
interaction between the components of the cosmic mixture are useful
to handle the coincidence problem, in the present study the
coincidence can not be evaded due to the choice of the solution
generating ansatz.
\end{abstract}

\maketitle

\section{Introduction}

At the time, all the available observational evidence suggest that
the energy density of the universe today is dominated by a component
with negative pressure of, almost, the same absolute value as its
energy density. This mysterious component, that violates the strong
energy condition (SEC), drives a present accelerated stage of the
cosmic evolution and is called "dark energy" (DE). A pleiad of
models have been investigated to account for this SEC violating
source of gravity. Among them, one of the most successful models is
a slowly varying scalar field, called
'quintessence'\cite{kolda,quintessence}.

Many models of quintessence assume that the background and the dark
energy evolve independently, so their natural generalization are
models with non-minimal coupling between both components. Although
experimental tests in the solar system impose severe restrictions on
the possibility of non-minimal coupling between the DE and ordinary
matter fluids \cite{will}, due to the unknown nature of the dark
matter (DM), it is possible to have additional (non gravitational)
interactions between the DE and the DM components, without conflict
with the experimental data.\footnotemark\footnotetext{It should be
pointed out, however, that when the stability of dark energy
potentials in quintessence models is considered, the coupling dark
matter-dark energy is troublesome \cite{doram}.} Since, models with
non-minimal coupling imply interaction (exchange of energy) among
the DM and the DE, these models provide new qualitative features for
the coincidence problem \cite{luca,pavon}. It has been shown, in
particular, that a suitable coupling, can produce scaling solutions
that are free of the coincidence. The way in which the coupling is
approached is not unique. In reference \cite{luca}, for instance,
the coupling is introduced by hand. In \cite{pavon} the type of
coupling is not specified from the beginning. Instead, the form of
the interaction term is fixed by the requirement that the ratio of
the energy densities of DM and quintessence had a stable (non-zero)
equilibrium point that solves the problem of the coincidence. In
\cite{chimento}, a suitable interaction between the quintessence
field and DM leads to a transition from the matter domination era to
an accelerated expansion epoch in the model of Ref.\cite{pavon}. A
model derived from the Dilaton is studied in \cite{bean1}. In this
model the coupling function is chosen as a Fourier expansion around
some minimum of the scalar field.

A variety of self-interaction potentials have been studied in DE
models to account for the desired evolution that fits the
observational evidence. Among them, a single exponential is the
simplest case. This type of potential leads to two possible
late-time attractors in the presence of a barotropic
fluid\cite{wands,copeland}: i) a scaling regime where the scalar
field mimics the dynamics of the background fluid, i. e., the ratio
between both DM and quintessence energy densities is a constant and
ii) an attractor solution dominated by the scalar field. Given that
single exponential potentials can lead to one of the above scaling
solutions, then it should follow that a combination of exponentials
should allow for a scenario where the universe can evolve through a
radiation-matter regime (attractor i)) and, at some recent epoch,
evolve into the scalar field dominated regime (attractor ii)).
Models with single and double exponential potentials has been
studied also in references \cite{ruba,Isra1}.

Aim of the present paper is to investigate models with non-minimal
coupling among the components of the cosmic fluid: the dust dark
matter and the quintessence field (a scalar field model of dark
energy). To specify the kind of coupling, we take as a Lagrangian
model, a scalar-tensor (ST) theory with Lagrangian written in
Einstein frame variables. Otherwise, one might also consider the
Lagrangian with additional non-gravitational coupling between the
matter species as an effective theory.

In this paper, to derive solutions of the field equations of the
model, we use an ansatz that makes possible easy handling of the
differential equations involved at the cost, however, of loosing the
possibility to avoid the coincidence problem through an appropriate
choice of the interaction between the DM and the DE (quintessence
field). Actually, we impose the dynamics of the expansion by
exploring a linear relationship between the Hubble parameter and the
time derivative of the scalar field.\footnotemark\footnotetext{This
relationship is the simplest possible and, in the absence of any
other information is, in a sense, the most natural since the Hubble
parameter fixes the time scale. This argument was suggested to us by
Diego Pavon.} Using this relationship we can solve the field
equations explicitly. However, as it will be shown, unlike other
models where the coincidence problem is solved (or smoothed out)
through the choice of an appropriated interaction between the DM and
the quintessence field, in the present investigation, solutions
where the ratio between the dark matter and the quintessence energy
densities is a constant $\sim 1$ (it is the necessary requirement to
avoid the coincidence problem), are unstable so that the coincidence
do arises.

Here we consider a flat Friedmann-Robertson-Walker (FRW) universe
filled with a mixture of two interacting fluids: a background of DM
and the quintessence field. Since there are suggestive arguments
showing that observational bounds on the "fifth" force do not
necessarily imply decoupling of baryons from
quintessence\cite{pavonx}, then baryons are to be considered also as
part of the background DM interacting with the quintessence field.
In other words, like in reference \cite{pavonx} we are considering
universal coupling of the quintessence field to all sorts of matter
(radiation is excluded). Since the arguments given in the appendix
of reference \cite{pavonx} to explain this possibility are also
applicable in the cases of interest in the present study, we refer
the interested reader to that reference to look for the details.
However we want to mention the basic arguments given therein: A
possible explanation is through the "longitudinal coupling" approach
to inhomogeneous perturbations of the model. The longitudinal
coupling involves energy transfer between matter and quintessence
with no momentum transfer to matter, so that no anomalous
acceleration arises. In consequence, this choice is not affected by
observational bounds on "fifth" force exerted on the baryons. Other
generalizations of the given approach could be considered that do
involve an anomalous acceleration of the background due to its
coupling to quintessence. However, due to the universal nature of
the coupling, it could not be detected by differential acceleration
experiments. Another argument given in \cite{pavonx} is that, since
the coupling chosen is of phenomenological nature and its validity
is restricted to cosmological scales (it depends on magnitudes that
are only well defined in that setting), the form of the coupling at
smaller scales remains unspecified. The requirements for the
different couplings that could have a manifestation at these scales
are that a) they give the same averaged coupling at cosmological
scales, and b) they meet the observational bounds from the local
experiments. We complete these arguments by noting that they are
applicable even if the coupling is not of phenomenological origin,
like in the present investigation where the kind of coupling chosen
is originated in a scalar-tensor theory of gravity.

The paper has been organized as follow. In section II the details of
the model are given. The method used to derive FRW solutions is
explained in section III. We use the 'two fluids' approach: a
background fluid of DM and a self-interacting scalar field
(quintessence). Specific exponential couplings with different
exponents, that are inspired in ST theory and lead, correspondingly,
to self-interaction potentials of single exponential and double
exponential class, are studied separately. In section IV a study of
existence and stability of the solutions is presented. We point out
that, due to the choice of the solution generating ansatz, the
coincidence problem can not be evaded since, a constant ratio
between the energy densities of the components of the cosmic mixture
is never a stable attractor. In section V, conclusions are given.

\section{The model}

We consider the following action that is inspired in a scalar-tensor
theory written in the Einstein frame, where the quintessence
(scalar) field is coupled to the matter degrees of freedom:

\begin{eqnarray}
S=\int_{M_4}d^4x\sqrt{|g|}\{\frac{1}{2}R-\frac{1}{2}(\nabla\phi)^2-V(\phi)
\nonumber\\+ C^2(\phi){\cal L}_{(matter)}\}.\label{action}
\end{eqnarray}
In this equation $R$ is the curvature scalar, $\phi$ is the scalar
(quintessence) field, $V(\phi)$ - the quintessence self-interaction
potential, $C^2(\phi)$ - the quintessence-matter coupling function,
and ${\cal L}_{matter}$ is the Lagrangian density for the matter
degrees of freedom. This action could be considered, instead, as an
effective theory, implying additional non-gravitational interaction
between the components of the cosmic fluid.

Although in the present study we will be concerned mainly with FRW
spacetimes with flat spatial sections, here for generality we write
the FRW line element in the form:

\bea &&ds^2=-dt^2+a(t)^2(\frac{dr^2}{1-kr^2}+r^2
d\Omega^2),\nonumber\\&&d\Omega^2\equiv d\theta^2+\sin^2\theta
d\phi^2,\eea where $t$ is the cosmic time, $r,\theta,\phi$ are the
spatial (radial and angular) coordinates and $k$ is the spatial
curvature, which we take to be zero in this investigation. We use
the system of units in which $8\pi G=c=\hbar=1$.

The spacetime is filled with a background pressureless dark matter
fluid and a quintessence field (the scalar field $\phi$). As already
said the baryons (a subdominant component at present, but important
in the past of the cosmic evolution) are included in the background
of dark matter. In the introductory part of this paper we have
already commented on the possibility of a universal coupling of the
dark energy to all sorts of matter, including the baryons (and
excluding radiation).

The field equations that are derived from the action (\ref{action})
are:

\begin{eqnarray}
3H^{2}+\frac{3k}{a^2}&=&\rho_{m}+\frac{1}{2}\dot\phi^{2}+V\label{frieman1},\\
2\dot H+3H^{2}+\frac{k}{a^2}&=&(1-\gamma)\rho_{m}-
\frac{1}{2}\dot\phi^{2}+V \label{frieman2},\\
\ddot\phi+3H\dot\phi &=&-V'+(\ln{X})'\rho_{m} \label{frieman3},
\end{eqnarray}
where we have introduced the reduced notation $X(\phi)\equiv
C(\phi)^{(3\gamma-4)/2}$. The parameter $\gamma$ is the barotropic
index of the background fluid (DM). The "continuity" equation for
the background is:
\begin{eqnarray} \dot\rho_{m}+3\gamma H
\rho_{m}=-(\ln{X})'\dot\phi\;\rho_{m} \label{ecuaciondinamica1},
\end{eqnarray} or, after integration
\begin{eqnarray}
\rho_{m}=Ma^{-3\gamma}X^{-1}\label{ecuaciondinamica2}.\end{eqnarray}
where $M$ is a constant of integration. In the former equations the
dot accounts for derivative in respect to the co-moving time $t$,
while the prime denotes derivative in respect to $\phi$. We now
proceed to derive exact solutions to the above field equations by
fixing the dynamics of the expansion.

\section{Deriving Solutions}

In order to derive exact analytic (flat; $k=0$) solutions, one
should either fix the dynamics of the cosmic evolution or fix the
functional form of the self-interaction potential $V(\phi)$. In the
present study we fix the dynamics of the model by imposing the
following constraint
\begin{eqnarray} \dot\phi=\lambda H,
\label{restriccion1}\end{eqnarray} that involves the Hubble
parameter and the square root of the kinetic energy of the
quintessence field ($\lambda$ is an arbitrary constant parameter).
As it will be immediately shown, this relationship enables one to
reduce the system of differential equations
(\ref{frieman1}-\ref{frieman3}) to a single first order differential
equation, involving the self-interaction potential $V$ together with
its derivative in respect to the scalar field variable $\phi$ and
the corresponding derivative of the coupling function. Therefore if
one chooses further the form of the coupling function, the
functional form of the self-interaction potential can be found by
solving of the corresponding differential equation. As it will be
shown, exponential coupling functions with different exponents
represent the simplest situations to study in the present model.
Correspondingly, the ansatz (\ref{restriccion1}) implies that only
self-interaction potentials of the exponential form (including their
combination) can be considered. Integration of equation
(\ref{restriccion1}) implies that
\begin{eqnarray} a=e^{\phi/\lambda},\label{restriccion}\end{eqnarray}
where the scale factor has been normalized so that the integration
constant has been absorbed into it. After this, the equation
(\ref{ecuaciondinamica2}) can be rewritten in the following form:
\begin{eqnarray}
\rho_{m}(\phi)=Me^{-\frac{3\gamma}{\lambda}\phi}X^{-1}(\phi).
\label{expresionro}\end{eqnarray} If one adds up equations
(\ref{frieman1}) and (\ref{frieman2}) then one obtains:
\begin{eqnarray} \dot H+3H^{2}=\frac{2-\gamma}{2}\rho_{m}+V.
\label{sumafrieman}\end{eqnarray} Substituting (\ref{restriccion1})
in (\ref{frieman3}), and comparing the resulting equation with
(\ref{sumafrieman}), a differential equation relating $V$ and the
coupling function $X$ (and their derivatives in respect to the
scalar field variable $\phi$) can be obtained:
\begin{eqnarray} \frac{dV}{d\phi}+\lambda
V(\phi)=\rho_{m}(\phi)\left(\frac{d\ln{X}}
{d\phi}-\frac{\lambda(\gamma-2)}{2}\right).\label{equacionvx}\end{eqnarray}

Consider further equation (\ref{restriccion1}) written in the form
$d\phi=\lambda d(\ln a)$. It is then worthwhile rewriting of the
equation (\ref{equacionvx}) in the following form:

\begin{eqnarray}
V'+\lambda^{2}V=\left(\frac{X'}{X}-\frac{\lambda(\gamma-2)}{2}
\right)\frac{M}{a^{3\gamma}}X^{-1},\label{equacionvx2}
\end{eqnarray}
where now the prime denotes derivative in respect to the variable
$N=\ln a$ and we have substituted $\rho_{m}(\phi)$ from equation
(\ref{expresionro}). Equation (\ref{frieman1}) can then be
integrated in quadratures:

\begin{eqnarray}
\int \frac{d \phi}{\sqrt{M
e^{-\frac{3\gamma}{\lambda}\phi}X^{-1}(\phi)+V(\phi)}}=\sqrt{\frac{2
\lambda^{2}}{6-\lambda^{2}}}(t+t_{0}),\label{exp14}
\end{eqnarray}
or, if one introduces the time variable $d\tau
=e^{-\frac{3\gamma}{2\lambda}\phi}X^{-1/2}dt$,

\begin{eqnarray}
\int \frac{d\phi}{\sqrt{M
+e^{\frac{3\gamma}{\lambda}\phi}X(\phi)V(\phi)}}= \sqrt{\frac{2
\lambda^{2}}{6-\lambda^{2}}} (\tau+\tau_{0}). \label{exp16}
\end{eqnarray}

In consequence, once the function $X(\phi)$ (or $X(a)$) is given as
input, then one can solve equation (\ref{equacionvx}) (or
(\ref{equacionvx2})) to find the functional form of the potential
$V(\phi)$ (or $V(a)$). The integral (\ref{exp14}) (or (\ref{exp16}))
can then be taken explicitly to obtain $t=t(\phi)$ (or
$\tau=\tau(\phi)$). By inversion we can obtain $\phi=\phi(t)$ (or
$\phi=\phi(\tau)$) so, the scale factor can be given as function of
either the cosmic time $t$ or the time variable $\tau$ through
Eqn.(\ref{restriccion}).

\subsection{Particular Cases}

We shall study separately the simplest situations that can be
considered once the choice (\ref{restriccion1}) is made. In both
cases one deals with coupling functions of the exponential form:
$X(\phi)=X_0 \exp{(n\phi)}$, where $X_0$ and $n$ are constant
parameters. If one chooses $n=(2-3\gamma/2)/\sqrt{\omega+3/2}$ where
$\omega$ is the Brans-Dicke coupling parameter, then the action
(\ref{action}) corresponds to Brans-Dicke theory written in the
Einstein frame (EF). In this case the EF scalar field $\phi$ is
related to the Jordan frame scalar field $\hat\phi$ through:
$d\phi=d\hat\phi/(\omega+3/2)$. In general, once the dynamics
(\ref{restriccion1}) is imposed, this kind of coupling function
leads to double and single exponential potentials depending on $r$.
The importance of this class of potentials in cosmology has been
already outlined in the introductory part of this paper.

\subsubsection{Case A}

Let us consider the simplest situation when, in Eq.
(\ref{equacionvx});
\begin{eqnarray}
\frac{d\ln X}{d\phi}=\frac{\lambda(\gamma-2)}{2}\label{xenB},
\end{eqnarray} i. e., we are faced with an exponential coupling
function of the form mentioned above with $r=\lambda(\gamma-2)/2$ in
the exponent. In this case the equation (\ref{equacionvx})
simplifies to

\begin{eqnarray}
\frac{d V}{d \phi} + \lambda V =0, \label{exp34A}
\end{eqnarray}
which can be easily integrated to yield to a single exponential
potential:
\begin{eqnarray}
V=V_{0} e^{-\lambda \phi}. \label{potenB}
\end{eqnarray}
In consequence the equation (\ref{exp16}) can be written as

\begin{eqnarray}
\int \frac{d w}{\sqrt{w^{2}+A^{2}}} = \mu
(\tau+\tau_{0})\label{expr16enB}
\end{eqnarray}
where $w=\exp(-\frac{\gamma (6-\lambda^{2})}{4 \lambda}\phi)$,
$A^{2}=V_{0}X_{0}/M$ and $\mu=\gamma \sqrt{M (\lambda^{2}-6)/8}$, so
we have the explicit solution

\begin{eqnarray}
\phi(\tau)= \phi_{0} + \ln \left( \sinh \left[ \mu
(\tau+\tau_{0})\right]^{\frac{4\lambda}{\gamma (6-\lambda^{2})}}
\right), \label{exprfienB}
\end{eqnarray}
and, consequently:
\begin{eqnarray}
a(\tau)= a_{0} \sinh \left[ \mu
(\tau+\tau_{0})\right]^{\frac{4\lambda}{\gamma (6-\lambda^{2})}}.
\label{expraenB}
\end{eqnarray}

The dimensionless density parameter (for the i-th component
$\Omega_i=\rho_i/3H^2$) and the Hubble expansion parameter can be
given as functions of the redshift $z$ also. Actually, if one
considers that $a(z)=a_{0}/(1+z)$, where $a_0\equiv a(z=0)$ (for
simplicity of the calculations we choose the normalization
$a_{0}=1$) then;

\begin{eqnarray}
\Omega_{m}(z)=\frac{6-\lambda^{2}}{6 A^{2}} \frac{(1+z)^{3
\gamma+\gamma (6-\lambda^{2})/2 - \lambda^{2}}}{(1+z)^{3
\gamma+\gamma (6-\lambda^{2})/2 - \lambda^{2}}+A^{2}},
\label{omegaB}
\end{eqnarray}
and
\begin{eqnarray}
H(z)=B\sqrt{\frac{1}{A^{2}}(1+z)^{3 \gamma+\gamma
(6-\lambda^{2})/2}+(1+z)^{\lambda^{2}}},\label{hubbleB}
\end{eqnarray}
where $B=\sqrt{2 V_{0}/(6-\lambda^{2})}$. Note that, $\Omega_{m}$ is
a maximum at $z=\infty$: $\Omega_m(\infty)=(6-\lambda^2)/6A^2$. In
general one can write that, at the epoch of nucleosynthesis
$\Omega_{m}(\infty)=(1-\epsilon)$, where $\epsilon$ is a very small
number (the small fraction of dark energy component during
nucleosynthesis epoch) $\epsilon=[6(A^2-1)+\lambda^2]/6A^2$. Taking
into account the observational fact that, according to
model-independent analysis of SNIa data\cite{peebles}, at present,
($z=0$); $\Omega_{m}(0)=1/3$, then (\ref{omegaB}) can be rewritten
as

\begin{eqnarray}
\Omega_{m}(z)=\frac{(1-\epsilon)(1+z)^{3 \gamma+\gamma
(6-\lambda^{2})/2 - \lambda^{2}}}{(1+z)^{3 \gamma+\gamma
(6-\lambda^{2})/2 - \lambda^{2}}+(2-3\epsilon)}, \label{omegaBB}
\end{eqnarray} where now the solution exhibits only two free parameters
$\lambda$ and $\epsilon$ (the DM EOS parameter $\gamma=1$).

Other physical magnitudes of observational interest are the
quintessence equation of state (EOS) parameter and the deceleration
parameter, that are given by the following expressions:

\begin{eqnarray}
\omega_{\phi}=-1+\frac{\lambda^{2}}{3\Omega_\phi}, \label{expestado}
\end{eqnarray}
\begin{eqnarray}
q=-1+\frac{\lambda^{2}}{2}+\frac{3 \gamma}{2}\Omega_{m},
\label{expedesa}
\end{eqnarray}
respectively.

\begin{figure}[t!]
\begin{center}
\hspace{0.4cm}
\includegraphics[width=6.5cm,height=5cm]{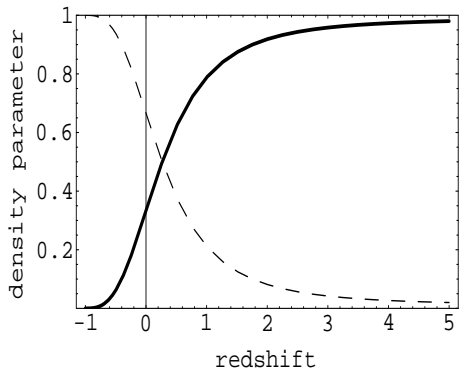}
\caption{The evolution of $\Omega_m$ (thick solid line) and
$\Omega_\phi$ (dashed line) vs $z$ is shown for the model with a
single exponential potential. The following values of the free
parameters $\varepsilon=0.01$ and $\lambda=0.3$ have been chosen.
Equality of matter and quintessence energy density occurs
approximately at $z\approx 0.3-0.4$. } \label{densi}
\end{center}\end{figure}

\begin{figure}[t!]
\begin{center}
\hspace{0.4cm}
\includegraphics[width=6.5cm,height=5cm]{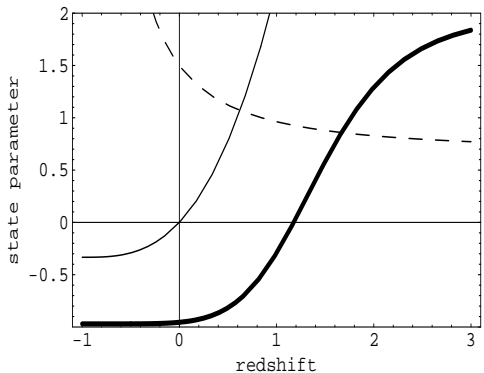}
\caption{We plot the dynamical EOS parameter of the quintessence vs
$z$, for the Model A (single exponential potential), for three
different values of the parameter $\lambda$: $0.3$ (thick solid
line), $1.41$ (solid line), and $2.24$ (dashed line) respectively.
In all cases the free parameter $\varepsilon$ is chosen to be
$\varepsilon=0.01$. Note that only the curve with the smallest value
of the parameter $\lambda$ (thick solid line) meets the requirements
of the present observational data favoring a value of the EOS
parameter $\omega_\phi\sim -1$.} \label{stateA}
\end{center}\end{figure}

\begin{figure}[t!]
\begin{center}
\hspace{0.4cm}
\includegraphics[width=6.5cm,height=5cm]{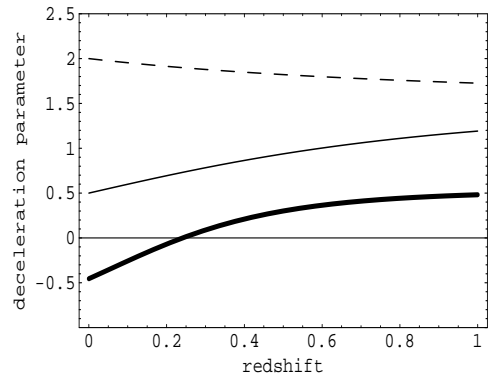}
\caption{We plot the evolution of the deceleration parameter $q$ vs
$z$ for the Model A (single exponential potential). As before, the
following values of the free parameters $\varepsilon=0.01$ and
$\lambda=0.3$ have been chosen. Again, the curve with the smallest
value of the parameter $\lambda$ is the one that fits better in the
observational data set on a present stage of accelerated expansion.}
\label{desA}
\end{center}\end{figure}

\subsubsection{Case B}

A second very simple choice is to consider, in Eq.
(\ref{equacionvx});
\begin{eqnarray}
\frac{d}{d\phi}\left(\ln{X(\phi)}\right)=const=\alpha\Rightarrow
X(\phi)=X_{0}e^{\alpha\phi},\label{acoplamientose}\end{eqnarray}
where $X_{0}$ is an integration constant. In this case, by
integration of (\ref{equacionvx}) one obtains
\begin{eqnarray}
V(\phi)=V_{0}e^{-\lambda\phi}+W_{0}e^{-(\alpha+3\gamma/\lambda)\phi},
\label{potencialdoble}\end{eqnarray} where the constant
\begin{eqnarray}
W_{0}=\frac{M}{2 X_{0}}\left(\frac{\lambda
(2-\gamma)-2\alpha}{\alpha+3\gamma/\lambda-\lambda}\right).
\label{exp18}
\end{eqnarray}
We are faced with a self-interaction potential that is a combination
of exponentials with different constants in the exponent. The
usefulness of this kind of potential has been already explained in
the introduction and will be briefly discussed later within the
frame of the present model, when we study the stability of the
corresponding solution. If one introduces the variable
$y=e^{-\frac{l}{2}\phi}$, the equation (\ref{exp16}) can be written
as

\begin{eqnarray}
\int \frac{dy}{\sqrt{y^{2}+b^{2}}}= -\frac{l}{2}\sqrt{\frac{2
\lambda^{2}}{6-\lambda^{2}}\frac{X_{0}V_{0}}{b^{2}}} (\tau+\tau_{0})
\label{exp20}
\end{eqnarray}
where $b^{2}=(X_{0}V_{0})/(M+W_{0}X_{0})$ and
$l=\alpha+3\gamma/\lambda-\lambda$. Integration of (\ref{exp20})
yields to:

\begin{eqnarray}
\phi(\tau)=\phi_{0}+ln\left(\sinh\left[\mu(\tau+\tau_{0})\right]\right)^{-2/l},
\label{exp22}
\end{eqnarray}
and, consequently:
\begin{eqnarray}
a(\tau)=a_{0}\left(\sinh\left[\mu(\tau+\tau_{0})\right]\right)^{-2/l}.
\label{exp23}
\end{eqnarray}
where $\mu=-\frac{l}{2}\sqrt{\frac{2\lambda^{2}}{6-\lambda^{2}}
\frac{X_{0}V_{0}}{a^{2}}}$. On the other hand, by using the
Friedmann equation (\ref{frieman1}) and inserting the potential
(\ref{potencialdoble}), one obtains the Hubble parameter as function
of the quintessence field $\phi$:

\begin{eqnarray}
H^{2}(\phi)=\frac{2 V_{0}}{6-\lambda^{2}} \left[ b^{2} e^{-(\alpha+3
\gamma / \lambda) \phi} +e^{-\lambda \phi} \right], \label{exp30A}
\end{eqnarray}
where we have considered Eqn.(\ref{expresionro}). In terms of the
scale factor one has, instead,
\begin{eqnarray}
H^{2}(a)=\frac{2 V_{0}}{6-\lambda^{2}} \left[ b^{2} a^{-(\alpha
\lambda+3 \gamma)} +a^{-\lambda^{2}} \right].\label{exp30}
\end{eqnarray}
Using the redshift variable $z$ the above magnitudes can be written
as follows:

\begin{eqnarray}
H^{2}(z)=\frac{2 V_{0}b^{2}}{6-\lambda^{2}} \left[(z+1)^{(\alpha
\lambda+3 \gamma )} +\frac{(1+z)^{\lambda^{2}}}{b^{2}}\right].
\label{exp31}
\end{eqnarray}
In turn, the energy density of DM can be written as
$\rho_{m}(z)=(M/X_{0})(z+1)^{\alpha \lambda +3 \gamma}=
\rho_{0}(z+1)^{\alpha \lambda +3 \gamma}$, so the DM dimensionless
density parameter can be given as function of $z$ also:
\begin{eqnarray}
\Omega_{m}(z)=\frac{(6-\lambda^{2}) \rho_{0}}{2 V_{0} b^{2}}
\frac{(z+1)^{\alpha \lambda +3 \gamma - \lambda^{2}}}{(z+1)^{\alpha
\lambda +3 \gamma - \lambda^{2}}+1/b^{2}}.\label{exp33}
\end{eqnarray}

In order to constrain the parameter space of the solution one should
consider the model to fit the observational evidence on a universe
with an early matter dominated period and a former transition to
dark energy dominance.\footnotemark\footnotetext{Since we are
dealing with models with non-minimal coupling between the
quintessence and the matter fields, here we refer, mainly, to a
model-independent analysis of SNIa observational
data\cite{peebles,observ,observq}.} Therefore $\lambda^{2}-\alpha
\lambda -3 \gamma<0$, besides, as before, one can write
$\Omega_{m}(\infty)=(1-\epsilon)$, where $\epsilon$ is a small
number (the small fraction of dark energy component during
nucleosynthesis epoch). In correspondence
$b^2=[(6-\lambda^{2})/(1-\epsilon)](\rho_{0}/2V_{0})$. Besides, if
one considers, as before, that according to model-independent
analysis of SNIa data\cite{peebles}, at present ($z=0$);
$\Omega_{m}(0)=1/3$, then $1/b^{2}=2-3\epsilon$. After this
Eqn.(\ref{exp33}) can be rewritten in the following way:

\begin{eqnarray}
\Omega_{m}(z)=(1-\epsilon) \frac{(z+1)^{\alpha \lambda +3 \gamma -
\lambda^{2}}}{(z+1)^{\alpha \lambda +3 \gamma - \lambda^{2}}+ (2-3
\epsilon)}, \label{exp33A}
\end{eqnarray} so the solution depends on three free parameters:
$\alpha$, $\lambda$ and $\epsilon$ (the barotropic index of the DM
is fixed: $\gamma=1$). These can be chosen so that the solution fits
well the observational data. Other physical magnitudes of
observational interest are the quintessence EOS parameter and the
deceleration parameter in equations (\ref{expestado}) and
(\ref{expedesa}), respectively.

In Figure \ref{densi}, we show the evolution of both dimensionless
DM and scalar-field energy densities $\Omega_{m}$ and
$\Omega_{\phi}$ respectively vs $z$ for the single exponential
potential (Case A). The corresponding evolution for Case B is very
similar. In Figure \ref{stateA}, the evolution of the dynamical
quintessence EOS parameter vs $z$ is plotted for the Model A (single
exponential potential), for three different values of the parameter
$\lambda$: $0.3$ (thick solid line), $1.41$ (solid line), and $2.24$
(dashed line) respectively. In all cases the free parameter
$\varepsilon$ is chosen to be $\varepsilon=0.01$. Note that only the
curve with the smallest value of the parameter $\lambda$ (thick
solid line) meets the requirements of the present observational data
favoring a value of the EOS parameter $\omega_\phi\sim -1$.
Meanwhile, in Figure \ref{desA}, we plot the evolution of the
deceleration parameter $q$ vs $z$ for the Model A. Again, the curve
with the smallest value of the parameter $\lambda$ is the one that
fits better the observational data set on a present stage of
accelerated expansion\cite{turner,triess}.

\begin{figure}[t!]
\begin{center}
\hspace{0.4cm}
\includegraphics[width=6.5cm,height=5cm]{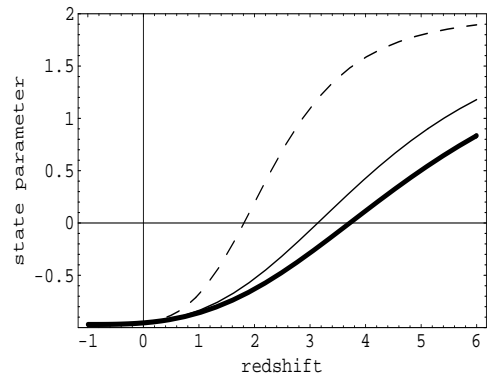}
\caption{We plot the dynamical EOS parameter of the quintessence vs
$z$, for the Model B (double exponential potential), for three
different values of the parameter $\alpha$: $0.1$ (thick solid
line), $1$ (solid line), and $5$ (dashed line) respectively. In all
cases the remaining free parameters are chosen to be
$\varepsilon=0.01$ and $\lambda=0.3$ respectively. At small redshift
the dependence upon $\alpha$ is only weak.} \label{stateB}
\end{center}\end{figure}

\begin{figure}[t!]
\begin{center}
\hspace{0.4cm}
\includegraphics[width=6.5cm,height=5cm]{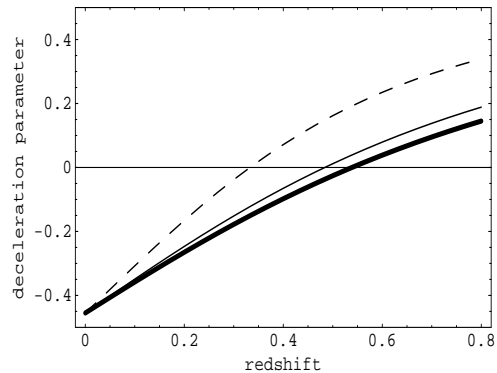}
\caption{We plot the evolution of the deceleration parameter $q$ vs
$z$ for the Model B, for three different values of the parameter
$\alpha$: $0.1$ (thick solid line), $1$ (solid line), and $5$
(dashed line) respectively. As before, the following values of the
free parameters $\varepsilon=0.01$ and $\lambda=0.3$ have been
chosen. The present value of the deceleration parameter is
independent on the value of $\alpha$.} \label{desB}
\end{center}\end{figure}

Figures \ref{stateB} and \ref{desB} show the behavior of
$\omega_\phi=\omega_\phi(z)$ and $q=q(z)$ for different values of
the free parameter $\alpha$: $0.1$ (thick solid line), $1$ (solid
line), and $5$ (dashed line). The free parameters $\varepsilon$ and
$\lambda$ have been fixed ad hoc, guided by the results of the study
of case A ($\varepsilon=0.01$, $\lambda=0.3$). It is apparent that
the present values of $\omega_\phi$ and of $q$, do not depend on
$\alpha$, so that this parameter can not be determined from the
observational data on the present stage of accelerated expansion of
the universe. This parameter could be of impact if the model is
applied to study early stages in the cosmic evolution.

\begin{table*}[t]
\begin{center}
\begin{tabular}{|c|c|c|c|}
$x$ & $y$ & Existence & Stability \\
\hline 1 & 1 & always & stable node if $0<\lambda<\sqrt{2}$; saddle
if
$\sqrt{2}<\lambda<\sqrt{6}$; unstable node otherwise\\
\hline $\frac{\lambda^{2}}{6}$ & 1 & $0<\lambda\leq\sqrt{6}$ & unstable node\\
\hline $\frac{\lambda^{2}}{2}$ & $\frac{\lambda^{2}}{2}$ & always &
saddle point if $0<\lambda<\sqrt{2}$; stable node if
$\lambda>\sqrt{2}$ \\
\end{tabular}
\end{center}
\caption[caseA1]{\label{caseA1} The properties of critical points
for case A.}
\begin{center}
\begin{tabular}{|c|c|c|}
$Point\;(x,y)$ & $\lambda_{1}$ & $\lambda_{2}$ \\
\hline $(1,1)$ & $\frac{\lambda^{2}-6}{2}$ & $\lambda^{2}-2$ \\
\hline $(\frac{\lambda^{2}}{6},1)$ & $3-\frac{\lambda^{2}}{2}$
& $\frac{\lambda^{2}+2}{2}$ \\
\hline $(\frac{\lambda^{2}}{2},\frac{\lambda^{2}}{2})$
& $-\frac{\lambda^{2}+2}{2}$ & $2-\lambda^{2}$ \\
\end{tabular}
\end{center}
\caption[caseA2]{\label{caseA2} The eigenvalues for Case A.}
\end{table*}

\begin{table*}[t]
\begin{center}
\begin{tabular}{|c|c|c|c|}
$x$ & $y$ & Existence & Stability \\
\hline 1 & 1 & always & stable node if $0<\lambda<\sqrt{2}$ and
$0<\alpha<\frac{3-\lambda^{2}}{\lambda}$; saddle point if either
$0<\lambda<\sqrt{2}$ and \\
 &  &  &  $\alpha<\frac{3-\lambda^{2}}{\lambda}$, or
$\sqrt{2}<\lambda<\sqrt{3}$ and
$0<\alpha<\frac{3-\lambda^{2}}{\lambda}$; unstable node otherwise \\
\hline $\frac{\lambda}{3}(\lambda+\alpha)$ & 1 &
$0<\lambda<\sqrt{3}$ & saddle point if $0<\lambda<\sqrt{2}$ and
$\frac{1}{\lambda}<\alpha<\frac{3-\lambda^{2}}{\lambda}$; \\
& & and
$0<\alpha\leq\frac{3-\lambda^{2}}{\lambda}$ & unstable node otherwise \\
\hline $\frac{\lambda^{2}}{2}$ & $\frac{\lambda^{2}}{2}$ & always &
stable node if $\lambda>\sqrt{2}$ and
$0<\alpha<\frac{1}{\lambda}$; unstable node if $0<\lambda<\sqrt{2}$ and  \\
& & & $\alpha>\frac{1}{\lambda}$; saddle point otherwise  \\
\end{tabular}
\end{center}
\caption[caseB1]{\label{caseB1} The properties of critical points
for case B.}
\begin{center}
\begin{tabular}{|c|c|c|}
$Point\;(x,y)$ & $\lambda_{1}$ & $\lambda_{2}$ \\
\hline $(1,1)$ & $\lambda^{2}-2$ & $\lambda^{2}+\lambda \alpha-3$ \\
\hline $(\lambda(\lambda+\alpha)/3,1)$ & $3-\lambda^{2}-\lambda
\alpha$ & $1-\lambda \alpha$ \\
\hline $(\frac{\lambda^{2}}{2},\frac{\lambda^{2}}{2})$ &
$\lambda \alpha-1$ & $2-\lambda^{2}$ \\
\end{tabular}
\end{center}
\caption[caseB2]{\label{caseB2} The eigenvalues for Case B.}
\end{table*}

\section{Existence and stability of the solutions}

We now turn to the study of the stability of the solutions found. In
order to keep the study as general as possible, we do not specify
any concrete model for the dark energy. We only require that the DE
EOS parameter $\omega_{de}\geq -1$. For this end we rewrite the
field equations (\ref{frieman1}-\ref{frieman3}) in the following
form. The Friedmann equation (we include the most general situation
with spatial curvature $k\neq 0$):
\begin{equation}
3H^2+3\frac{k}{a^2}=\rho_m+\rho_{\phi}, \label{feq1}
\end{equation}
the Raychaudhuri equation:
\begin{equation}
2\dot H-2\frac{k}{a^2}=-(p_m+\rho_m+p_{\phi}+\rho_{\phi}),
\label{feq2}
\end{equation}
the continuity equation for the quintessence field:
\begin{equation}
\dot\rho_{\phi}+3H(\rho_{\phi}+p_{\phi})=-Q, \label{feq3}
\end{equation}
and the continuity equation for the background matter:
\begin{equation}
\dot\rho_{m}+3H(\rho_{m}+p_{m})=Q, \label{feq4}
\end{equation}
where the dot accounts for derivative with respect to the cosmic
time and $Q=-(\ln X)'\rho_{m}$ is the interaction term. Through this
section the prime will denote derivative in respect to the new
variable $N=\ln a$, that is related with the cosmic time through
$dN=H dt$.

Let us introduce the following dimensionless phase space
variables:\footnotemark\footnotetext{See reference \cite{odintsov}
for an alternative treatment of a related stability study.}
\begin{equation}
x\equiv\Omega_{\phi},;\;\;y\equiv\Omega_{tot}=\Omega_{\phi}+\Omega_m.
\label{dimensionless}
\end{equation}
After this, the Friedmann equation (\ref{feq1}) can be rewritten in
the following way:

\begin{equation}
y=1+\frac{k}{a^2H^2}. \label{dimensionless1}
\end{equation}

The governing equations (\ref{feq1}),(\ref{feq2}),(\ref{feq3}), and
(\ref{feq4}) can be written in terms of the phase space variables in
the following way:
\begin{eqnarray}
x'&=&(y-x)(\ln\chi)'+x(y-1+3\omega_{\phi}(x-1))\nonumber\\
y'&=&(y-1)(y+3\omega_{\phi}x). \label{dsystem}
\end{eqnarray}
The above equations represent an autonomous system of equations if
$(\ln\chi)'$ and $\omega_\phi$ do not depend on $N$ explicitly. In
the remaining part of this section, for simplicity, we assume that
it is the case, so that the system (\ref{dsystem}) is an autonomous
one. Besides we consider $\chi(a)=\chi_0\;a^\delta$, where $\delta$
is some constant parameter. This choice of the coupling function
comprises many useful situations (including the solutions we have
derived before) and it implies that $(\ln\chi)'=\delta$. We assume,
also, that $\omega_m=0\Rightarrow\gamma=1$, i. e., the background
fluid is dust. For the flat space case ($k=0$), the system
(\ref{dsystem}) should be complemented with the following constraint
equation:

\be 0\leq y-x\leq 1,\label{constraint}\ee which follows from
requiring that the positive dimensionless matter energy density
parameter $\Omega_m\leq 1$.

The first step towards the study of the dynamics of the autonomous
system (\ref{dsystem}) is to find its critical points
$(x_c,y_c)\Rightarrow (x',y')=(0,0)$. Then one can investigate their
stability by expanding equations (\ref{dsystem}) in the vicinity of
the critical points $x=x_c+u$, $y=y_c+v$ (up to terms linear in the
perturbations $u,v$):

\be\left(\begin{array}{c} u' \\  v' \\
\end{array}\right)=\Lambda \left(\begin{array}{c} u \\  v \\
\end{array}\right),\ee where $\Lambda$ is the matrix of the
coefficients in the expansion. The general solution for the
evolution of the linear perturbations can be written as:

\bea u=u_1\;e^{\lambda_1 N}+u_2\;e^{\lambda_2 N},\nonumber\\
v=v_1\;e^{\lambda_1 N}+v_2\;e^{\lambda_2 N},\label{linearperts}\eea
where $\lambda_1$ and $\lambda_2$ are the eigenvalues of the matrix
$\Lambda$.

\begin{figure}[t!]
\begin{center}
\hspace{0.4cm}
\includegraphics[width=6.5cm,height=5cm]{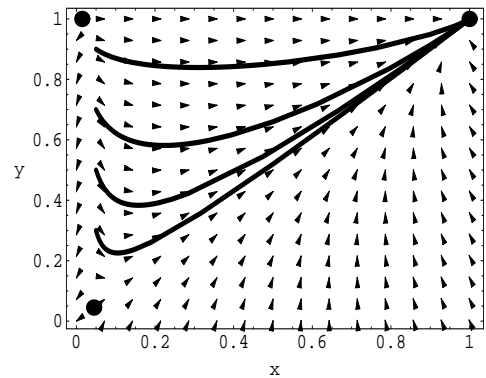}
\caption{The phase plane for Case A ($\gamma=1$, $\lambda=0.3$). The
critical point ($1$, $1$) is stable (a sink) so that the
quintessence dominated solution is the late time attractor. The
scaling regime (0.015, 1) is an unstable point. The saddle is
located at (0.045, 0.045). All the phase space trajectories diverge
from the unstable point and converge towards the attractor.}
\label{stabiA1}
\end{center}\end{figure}

In table \ref{caseA1} we show the properties of the critical points
(including existence and stability) for the case A above, meanwhile,
in table \ref{caseA2}, the corresponding eigenvalues are given. In
this case $X=X_0\;e^{-\lambda\phi/2}$. As seen, for the values of
$\lambda$ that make the model fit better the observational data set
($0\leq\lambda<1$), the first critical point
$(x,y)=(1,1)\Rightarrow\Omega_\phi=1$ (quintessence dominated phase)
is a stable attractor, while the scaling solution
$(x,y)=(\lambda^2/6,1)\Rightarrow\Omega_m/\Omega_\phi=6/\lambda^2-1>1$,
is always unstable. The phase portrait for this case is shown in
fig.\ref{stabiA1}. All of the trajectories in phase plane $(x,y)$,
diverge from the unstable point (matter dominated scaling solution)
and converge towards the attractor (quintessence dominated)
solution.

\begin{figure}[t!]
\begin{center}
\hspace{0.4cm}
\includegraphics[width=6.5cm,height=5cm]{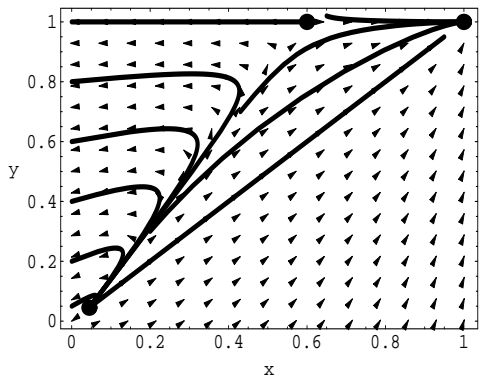}
\caption{The phase plane for Case B ($\gamma=1$, $\lambda=0.3$ and
$\alpha=5.7$). The quintessence dominated solution (point ($1$, $1$)
in the phase plane) is an attractor. The scaling solution (critical
point (0.6, 1)) is a saddle, meanwhile (0.045, 0.045) is an unstable
point. For different values of the parameter $\alpha$, the
separation between the saddle (scaling phase) and the stable
(quintessence dominated) critical points varies.} \label{stabiA2}
\end{center}\end{figure}

The properties of the critical points and the corresponding
eigenvalues for case B, where $X=X_0\;e^{\alpha\phi}$, are shown in
tables \ref{caseB1} and \ref{caseB2} respectively. It is apparent
that, for the relevant ranges of the free parameters
$0\leq\lambda<1$ and $0<\alpha<3/\lambda-\lambda$, the quintessence
dominated solution (first critical point $(x,y)=(1,1)$) is always a
stable node. The (matter dominated) scaling solution- second
critical point ($(x,y)=(\lambda(\lambda+\alpha)/3,1)$)- could be
either a saddle ($1/\lambda<\alpha<3/\lambda-\lambda$) or
,otherwise, an unstable node. The phase portrait fig.\ref{stabiA2}
shows that, for the values of the free parameters chosen
($\lambda=0.3$, $\alpha=5.7$), all phase space trajectories diverge
from the unstable node (third critical point
$(x,y)=(\lambda^2/2,\lambda^2/2)$-quintessence dominated solution
with curvature) and, either converge towards the stable attractor
solution dominated by the quintessence (first critical point), or
are repelled by the saddle point (second critical point- the matter
dominated scaling solution). This result, that is generic for both
solutions A and B, shows that, since the scaling (quintessence
dominated) solution with $\Omega_m/\Omega_\phi\lesssim 1$ is not
even a critical point of the corresponding autonomous dynamical
system (equations (\ref{dsystem})), then the coincidence problem
could arise in the cases studied in the present investigation. In
the next subsection we will show in a more definitive manner that
this is indeed the case.

\subsection{The coincidence problem}

Let us investigate whether, in the situations of interest in the
present study, the question: why are the energy densities of the
dark matter and of the dark energy of the same order precisely at
present? arises. For this purpose it is recommended to study the
dynamics of the ratio\cite{pavonx}:

\be
r=\frac{\rho_m}{\rho_\phi}=\frac{\Omega_m}{\Omega_\phi},\label{ratio}\ee
in respect to the variable $N\equiv \ln a$, that, as said before, is
related with the cosmic time $t$ through $dN=H dt$. The following
generic evolution equation holds for $r$:

\be r'=f(r),\label{evolutionr}\ee where the prime denotes derivative
in respect to the variable $N$ and $f$ is an arbitrary function (at
least of class ${\cal C}^1$) of $r$. One is then primarily
interested in the equilibrium points of equation (\ref{evolutionr}),
i.e., those points $r_{ei}$ at which $f(r_{ei})=0$. After that one
expands $f$ in the neighborhood of each equilibrium point;
$r=r_{ei}+\epsilon_i$, so that, up to terms linear in the
perturbations $\epsilon_i$: $f(r)=(df/dr)_{r_{ei}}\epsilon_i+{\cal
O}(\epsilon_i)\Rightarrow \epsilon_i'=(df/dr)_{r_{ei}}\epsilon_i$.
This last equation can be integrated to yield the evolution of the
perturbations:

\be
\epsilon_i=\epsilon_{0i}\exp{[(df/dr)_{r_{ei}}N]},\label{perts}\ee
where $\epsilon_{0i}$ are arbitrary integration constants. It is
seen from (\ref{perts}) that, only those perturbations for which:

\be (df/dr)_{r_{ei}}<0,\ee decay with the time variable $N$, and the
corresponding equilibrium point is stable. The coincidence is evaded
if the point $\rho_m/\rho_\phi=r_{ei}\lesssim 1$ is stable.

Now, if we take into account the conservation equations
(\ref{feq3}), and (\ref{feq4}), where $Q=-(\ln X)'\rho_{m}$, and
since $\Omega_\phi=1/(r+1)$, implying that (see equation
(\ref{expestado})) the quintessence EOS parameter $\omega_\phi$ can
be written as function of $r$: $\omega_\phi=-1+\lambda^2(r+1)/3$,
then, for the cases under study here, the function $f$ can be given
by the following expression:

\be f(r)=r\{(\lambda^2-\delta)(r+1)-3\},\label{function}\ee where
$\delta=(\ln X)'=n\lambda$. For the case A: $\delta=-\lambda^2/2$,
while for the case B: $\delta=\lambda \alpha$. It can be easily
checked that, in both cases the only stable equilibrium point is the
one for which the ratio $r_{e0}=0$, i.e., it is the quintessence
dominated solution, so that the coincidence can not be evaded.

\section{Conclusions}

We have found a new parametric class of exact cosmological scaling
solutions in a theory with general non-minimal coupling between the
components of the cosmic mixture: cold dark matter and dark energy
(the quintessence field). Baryons, although subdominant at present,
in the past of the cosmic evolution played a relevant role and could
be included as part of the dark matter component in the present
setup. There are suggestive arguments showing that observational
bounds on the "fifth" force do not necessarily imply decoupling of
baryons from quintessence\cite{pavonx}. To specify the general form
of the coupling we considered a scalar tensor theory of gravity
written in the Einstein frame. An alternative interpretation is to
consider it as an effective theory, implying additional
non-gravitational interaction between the components of the cosmic
fluid.

In order to derive exact flat FRW solutions to the model under
study, we have assumed a linear relationship between the Hubble
expansion parameter and the time derivative of the scalar field.
Mathematically this assumption allows to reduce the original system
of second order differential equations to a single, first order
differential equation, involving only the self-interaction potential
and the coupling function (together with their first derivatives in
respect to the scalar field variable). However, simplicity of the
mathematical handling is at the cost of retaining the problem of the
coincidence. In fact, the solution generating ansatz
(\ref{restriccion1}), implies that solutions where the ratio
$\rho_m/\rho_\phi=const\sim 1$ are not stable. In the cases studied,
only the quintessence dominated solution is stable. Anyway, the
assumed relationship between the square root of the scalar field
kinetic energy and the Hubble parameter is the simplest possible
and, in the absence of any other information is, in a sense, the
most natural since the Hubble parameter fixes the time scale.

We have concentrated our study in exponential class of coupling
functions. Brans-Dicke theory is a particular member in this class.
Exponential coupling functions can lead to self-interaction
potentials of the following class: A) single exponential potential
and B) double exponential potential. The stability and existence of
the solutions found have been also studied. For this purpose we have
applied a fairly general method in which one does not need to
specify any model for the dark energy. In both cases (case A and
case B), the dynamical system exhibits three critical points. For
the values of the free parameters that are allowed by the
observational data, the scalar field (quintessence) dominated
solution is always an attractor, meanwhile the scaling (matter
dominated) solution can be either an unstable node or a saddle
point.

We conclude that models with non-minimal coupling between the dark
energy and the dark matter are easy to handle mathematically if one
assumes a suitable dynamics. This is some times, at the cost of
retaining the problem of the coincidence. The present investigation
could be complemented by the study of different classes of
self-interaction potentials and/or the choice of suitable dynamics
allowing to evade the coincidence problem; one of the motivations to
study models with interaction between the components of the cosmic
mixture.

\section*{Acknowledgments}

We thank Diego Pavon for useful comments and the MES of Cuba by
financial support of this research.


\end{document}